\documentclass[letter]{aa}

\usepackage{graphicx}
\usepackage{txfonts}
\usepackage{color}
\usepackage[colorlinks=true, allcolors=blue]{hyperref}

\newcommand{\kms}{\ensuremath{\mathrm{km}\,\mathrm{s}^{-1}}}

\newcommand{\hi } {{\rm H}\,{\small\rm I}}
\begin{document}

   \title{Gas-rich ``ultra-diffuse'' galaxies are consistent with the baryonic Tully-Fisher relation and with Milgromian dynamics}
   
   \author{F. Lelli
          \inst{1}
          }

   \institute{INAF $-$ Arcetri Astrophysical Observatory, Viale Enrico Fermi 5, 50125, Florence, Italy}

   \date{Received 27/06/2024; accepted 12/07/2024}

 
  \abstract{Some gas-rich ``ultra-diffuse'' galaxies (UDGs), which are extreme examples of low surface brightness (LSB) dwarf galaxies, have been reported to lack dark matter and to be offset from the baryonic Tully-Fisher relation (BTFR). If confirmed, these UDGs would represent a serious challenge for both $\Lambda$CDM galaxy-formation models and Milgromian dynamics. Here I demonstrate that these conclusions are very dubious due to underestimated uncertainties on inclinations and/or distances. First, I show that UDGs are offset from the BTFR in the same way as usual face-on LSB dwarfs due to systematic biases at low inclinations. Next, I analyze the two UDGs with the best available rotation-curve data. The first  (AGC\,242019) is ideally inclined for kinematic studies; MOND can fit the observed rotation curve with a distance of 12.5$\pm$0.6 Mpc, which is consistent with Virgocentric flow models. The second UDG (AGC\,114905) is close to face-on, so not ideal for kinematic studies; MOND can fit the observed rotation curve with a distance of 68$\pm$13 Mpc and inclination of $15^{\circ}\pm2^{\circ}$, which are consistent with existing data. In particular, I show that the disk inclination is more uncertain than previously estimated due to significant asymmetries (lopsidedness) in the stellar distribution. In conclusion, there is no strong evidence that gas-rich UDGs and gas-rich LSB dwarfs are distinct galaxy populations with different dynamical properties; instead, UDGs seem to be a subset of LSB dwarf galaxies biased toward face-on systems.}


   \maketitle
%

\section{Introduction}\label{sec:intro}

Low surface brightness (LSB) galaxies have historically played a key role in testing both dark matter (DM) models and alternative theories, such as Milgromian dynamics (or MOdified Newtonian Dynamics, MOND). In the DM context, LSB galaxies have been found to be DM dominated down to small radii \citep{deBlok1997, McGaugh1998}, so they have been prime targets to measure the inner slopes of DM halos, leading to the cusp--core debate \citep[see][for a review]{deBlok2010}. In the MOND context, LSB galaxies were predicted \citep{Milgrom1983b} and subsequently observed \citep{McGaugh1998b} to have slowly rising rotation curves and to show large mass discrepancies. This occurs because LSB galaxies have low surface densities in both stars and gas, so their internal accelerations are always below the MOND acceleration scale $a_0=1.2 \times 10^{-10}$ m s$^{-2}$ \citep{Milgrom1983b}. For reviews on MOND, see \citet{Famaey2012}, \citet{Milgrom2014} and \citet{Banik2022}. For reviews on LSB dwarf galaxies, see \citet{Lelli2022} on gas dynamics and \citet{Battaglia2022} on stellar dynamics.

In recent years, the LSB regime has gained renewed interest thanks to new facilities that allow the detection of LSB galaxies well below the historical limit of $\sim$23-24 mag arcsec$^{-2}$ in the $B$-band \citep{McGaugh1996}, down to $\sim$27--28 mag arcsec$^{-2}$ \citep[e.g.,][]{Mihos2015, Munoz2015}. In particular, \citet{vanDokkum2015} introduced the class of so-called ``ultra-diffuse'' galaxies (UDGs), defined as systems with central surface brightness lower than 24.5 $g$ mag arcsec$^{-2}$ and effective radius larger than 1.5 kpc. The UDG definition is rather arbitrary;  UDGs and classical dwarfs indeed lie on the same structural relations (such as stellar mass vs effective radius and stellar mass vs color), with UDGs representing the extreme LSB end of the usual dwarf population \citep{Conselice2018, Chilingarian2019, Iodice2020, Marleau2021, Zoller2024}. Even though I think there is no strong need for the new UDG class, in the following I   use the term UDG to merely distinguish them from the historical population of LSB dwarf galaxies.

Similarly to typical LSB dwarf galaxies, gas-poor UDGs are abundant in galaxy clusters, while gas-rich UDGs are abundant in galaxy groups and in the field. The internal dynamics of gas-poor UDGs can be probed using stellar kinematics or globular clusters, but these approaches have intrinsic complications due to the noncollisional nature of the tracers, as well as observational difficulties due to the extreme LSB regime \citep[see, e.g., the debated measurements for NGC 1052-DF2 in][]{vanDokkum2018, Martin2018, Laporte2019, Emsellem2019, Danieli2019}. In gas-rich UDGs, instead, atomic hydrogen (\hi) forms rotating disks, so high-resolution \hi\ observations can be used to trace rotation curves and build full mass models, similarly to typical dwarf and spiral galaxies.

Some gas-rich UDGs have been reported to be offset from the baryonic Tully-Fisher relation (BTFR) and to be DM deficient \citep{Mancera2019, Mancera2020, Hu2023, DU2024}. These conclusions have been disputed by \citet{Brook2021}, \citet{Banik2022b}, and \citet{Sellwood2022}. The BTFR is a fundamental prediction of MOND \citep{Milgrom1983a}; if an isolated galaxy is robustly found to be off  this relation, the MOND paradigm is ruled out. If confirmed, the anomalous properties of UDGs will also have key implications for $\Lambda$CDM models of galaxy formation, possibly suggesting that some UDGs are old, DM-free, tidal dwarf galaxies \citep{Lelli2015, Silk2019, vanDokkum2022, Gray2023}.

Here I revisit the situation showing that the uncertainties in galaxy distance and disk inclination have been underappreciated. Taking these uncertainties into account in the same way as is usually done for typical galaxies \citep[e.g.,][]{Li2018}, gas-rich UDGs are consistent with both the BTFR and MOND. 

\begin{figure}
\includegraphics[width=0.475\textwidth]{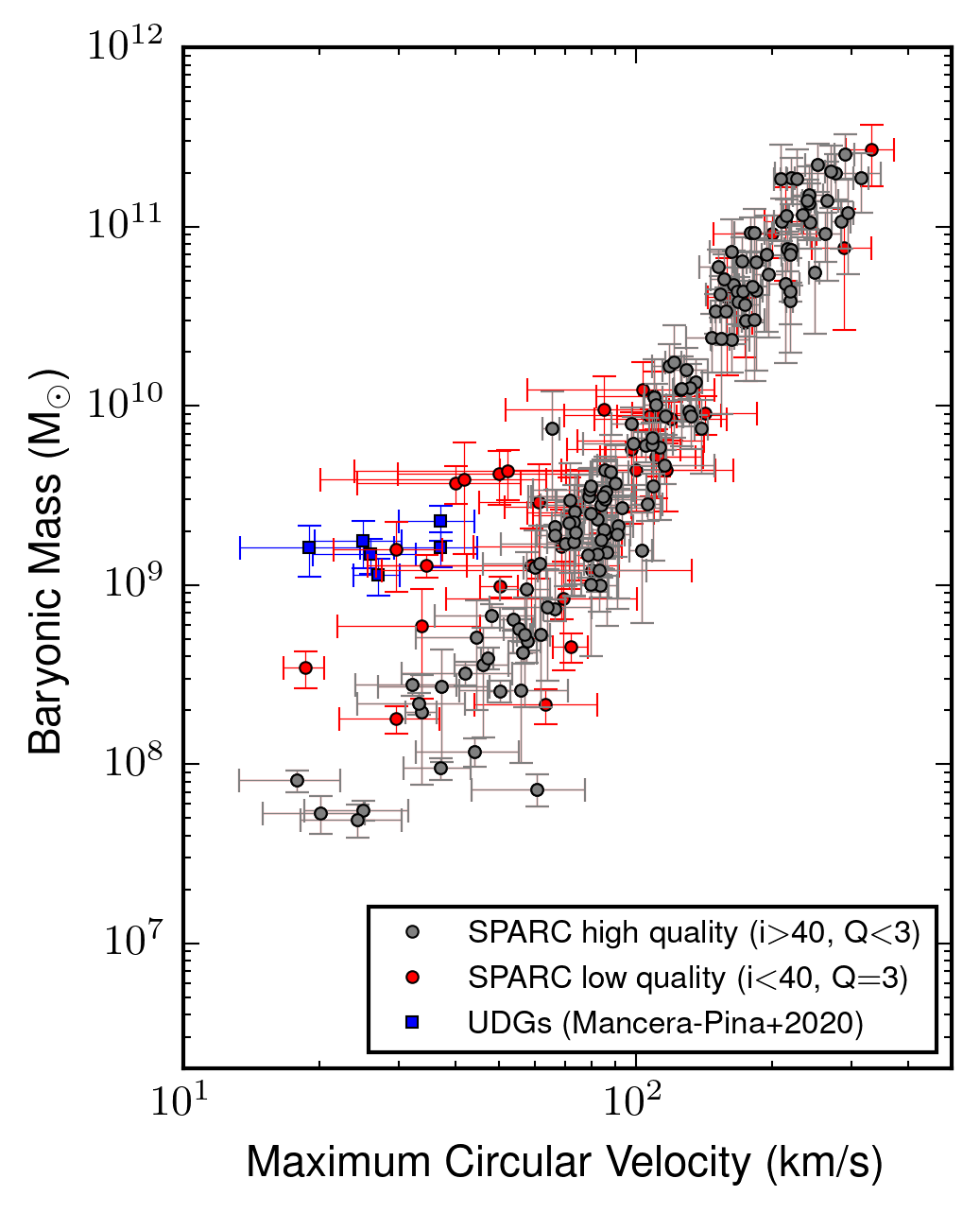}
\caption{Baryonic Tully-Fisher relation for high-quality SPARC galaxies (grey circles), low-quality SPARC galaxies (red circles), and UDGs (from \citealt{Mancera2020}; blue squares). Low-quality SPARC data and UDGs systematically scatter to the left of the BTFR, mostly due to inclination uncertainties in face-on disks. UDGs do not appear as a separate population when compared to LSB dwarfs with similarly low inclinations and data quality.}
\label{fig:BTFR}
\end{figure}

\section{Galaxy inclinations and the Tully-Fisher relation}\label{sec:inc}

When studying the Tully-Fisher relation and/or rotation curves, one generally avoids face-on galaxies with inclinations $i\lesssim40^{^\circ}$ \citep[e.g.,][]{Tully1977, Rubin1978, Sancisi1987, Begeman1991} for two basic reasons. First, the rotation velocities ($V_{\rm rot}$) are inferred by deprojecting the observed line-of-sight velocities ($V_{\rm los}$) using the formula $V_{\rm rot}\simeq V_{\rm l.o.s.}/\sin(i)$, so the term $1/\sin(i)$ becomes uncomfortably large for small inclination.  Second, it is extremely difficult to constrain inclinations below about 40$^{\circ}$ because the observed axial ratio $b/a \simeq \cos(i)$ varies  by a small amount, so symmetric disks with different $i$ appear nearly circular on the sky. In addition, actual galaxy disks are rarely symmetric in the outer regions (the lopsidedness phenomenon, \citealt{Baldwin1980, Richter1994}) and are often warped \citep[e.g.,][]{Sancisi1976, GarciaRuiz2002}, complicating substantially inclination estimates at low $i$. These facts are intrinsic limitations imposed to us by Nature; we can hardly do better by improving the data quality. 

Ultra-diffuse galaxies tend to be selected to be face-on systems because of another well-known fact. For an optically thin, dust-free stellar disk, the central observed surface brightness is given by $\mu_{\rm obs} = \mu_{\rm int} + 2.5\log[\cos(i)]$, where $\mu_{\rm int}$ is the intrinsic (face-on) surface brightness \citep[e.g.,][]{Tully1997}. When we select galaxies near the edge of the LSB detection limit of a survey, we tend to select face-on systems for which the term $2.5\log[\cos(i)]$ is small. For example, \citet{Mancera2020} discussed that the surface brightness distribution of their UDG sample is not what is  expected for randomly oriented disks on the sky, but biased toward low $i$. It is difficult, however, to determine how strong the inclination bias is because LSB disks are not entirely transparent (especially in optical bands), and different galaxies may have different levels of internal dust extinction. 

\begin{figure*}
\includegraphics[width=0.45\textwidth]{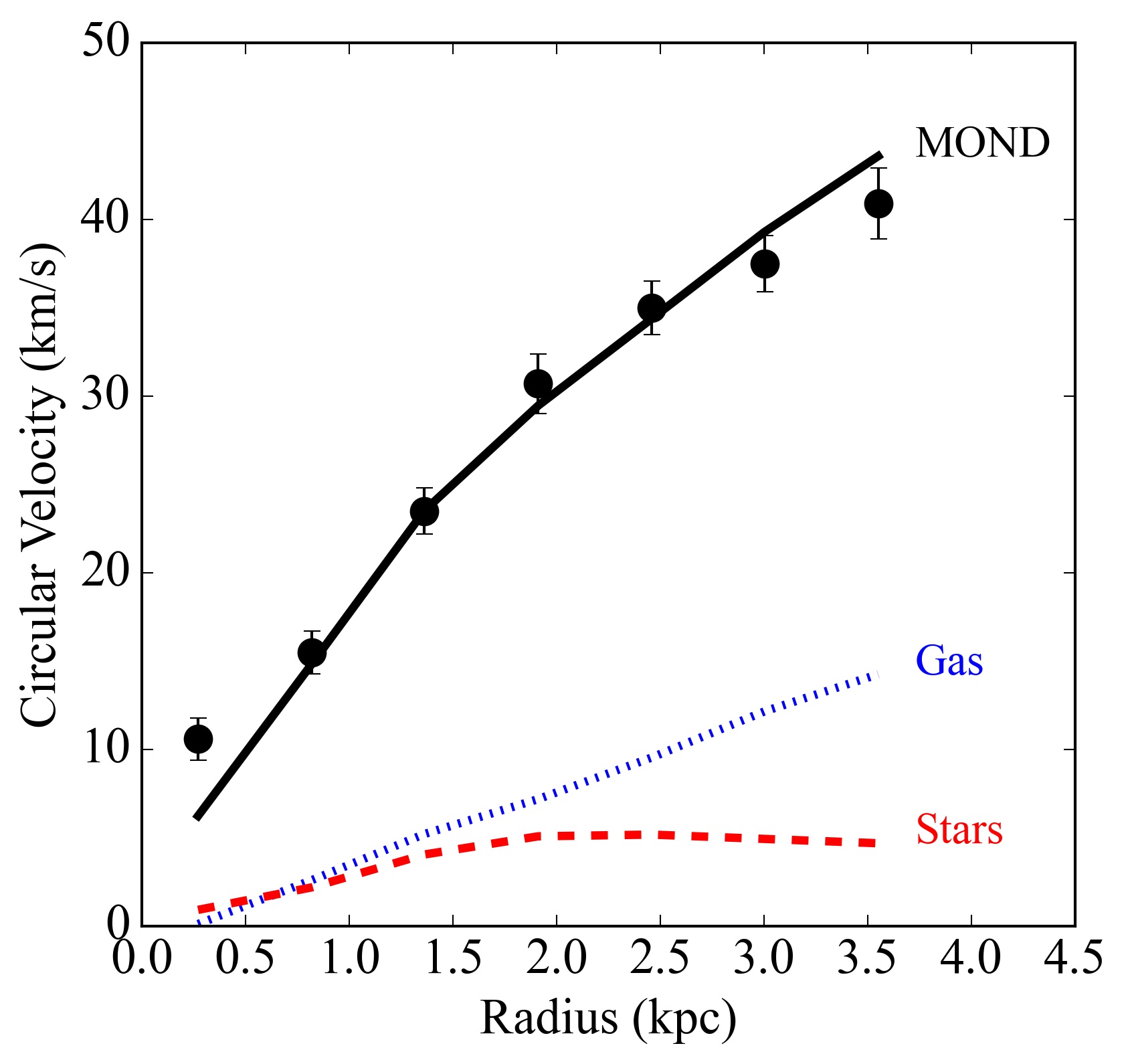}
\includegraphics[width=0.45\textwidth]{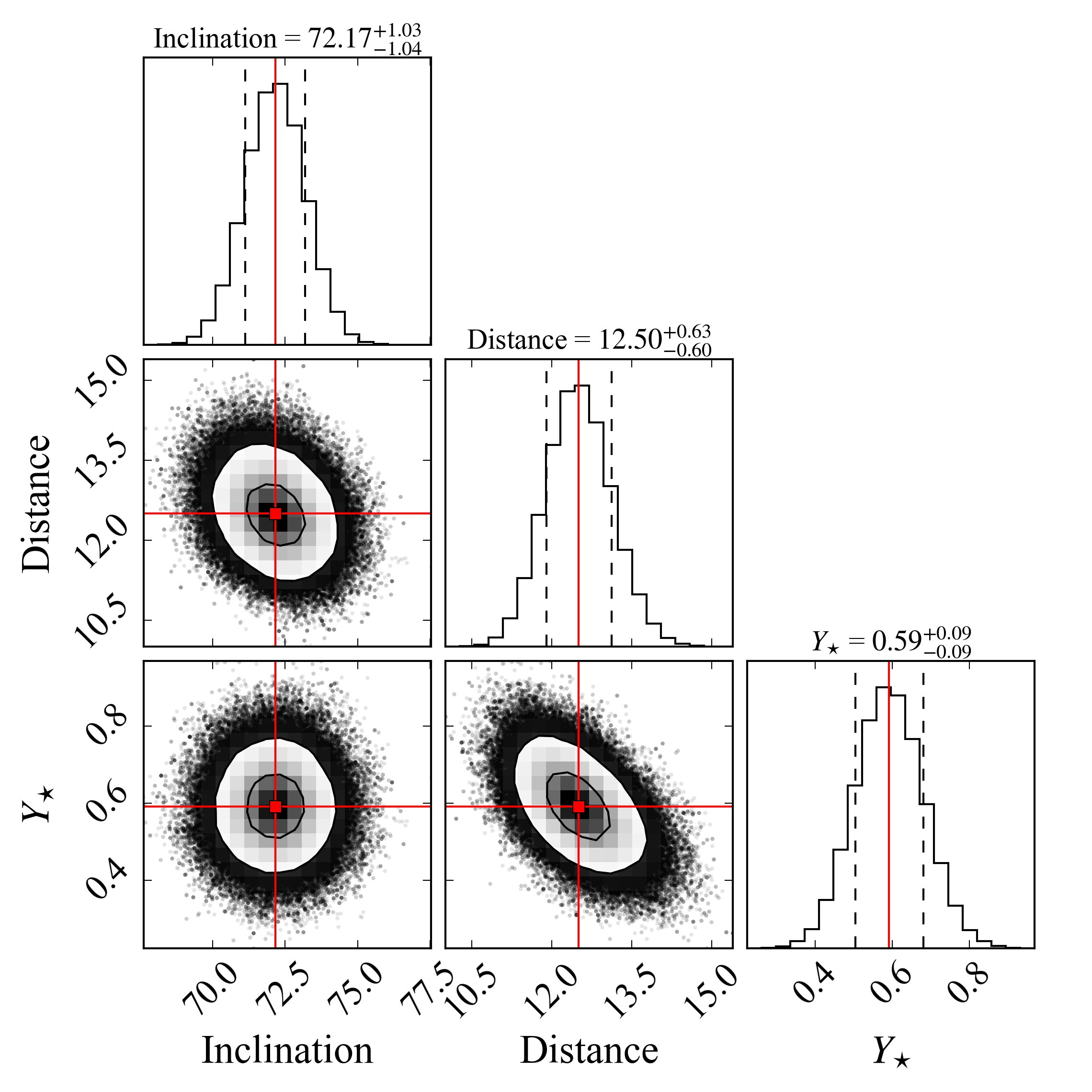}
\caption{Inclined ultra-diffuse galaxy AGC242019. \emph{Left panel}: Observed circular velocities (points with error bars) are well reproduced by MOND (solid black line) using the Newtonian contributions of stars (red dashed line) and gas (blue dotted line). \emph{Right panel}: Corner plot showing the posterior distribution of the nuisance parameters. MOND predicts a distance of 12.5 Mpc, which  is consistent with Virgocentric flow models.}
\label{fig:inclined}
\end{figure*}

Figure\,\ref{fig:BTFR} shows the BTFR from the Spitzer Photometry and Accurate Rotation Curve (SPARC) sample \citep{Lelli2016a}. To make a proper comparison with UDGs, I considered the maximum velocity along the rotation curve rather than the flat circular velocity because the \hi\ rotation curves of UDGs do not reach a clear flat part (see Figs. \ref{fig:inclined} and \ref{fig:faceon}). The use of the flat circular velocity minimizes the scatter in the BTFR \citep{Verheijen2001b, Lelli2019}, so it could give the artificial impression that UDGs are more offset from the BTFR than they really are.  Figure\,\ref{fig:BTFR} also includes low-quality SPARC galaxies with $i<40$ and/or $Q=3$, where $Q$ is a flag for strongly asymmetric systems \citep[see][]{Lelli2016a}. Low-quality galaxies systematically lie to the left of the mean relation. This is fully expected because disk asymmetries and lopsidedness tend to give systematically overestimated inclinations: a disk with $i\simeq0$ can be unambiguously recognized only if it is perfectly symmetric, which may never be the case in Nature. Notably, the SPARC sample contains only a small number of such low-quality galaxies because the \hi\ community historically pre-selected against face-on and asymmetric disks.

Gas-rich UDGs with spatially resolved data \citep{Mancera2020} scatter in the same direction as low-quality SPARC galaxies. UDGs do not look like a distinct population, considering that they have similarly low inclinations and data quality. The conclusions from \citet{Mancera2019, Mancera2020}, therefore, seem to be driven by improperly comparing face-on UDGs with top-quality SPARC galaxies, which have high inclinations, symmetric disks, and a well-defined flat rotation velocity, leading to a very tight BTFR \citep{Lelli2016b, Lelli2019}. In other words, the whole sample of \citet{Mancera2019, Mancera2020} would be excluded in BTFR studies when applying standard quality criteria. The same considerations apply to the conclusions of \citet{Hu2023} and \citet{DU2024}, who used \hi\ linewidths from spatially unresolved data. One must compare apples with apples; when doing so, UDGs are consistent with the BTFR of LSB dwarf galaxies with similar inclinations and data quality.

\section{The inclined galaxy AGC 242019}\label{sec:shi}

I now study the best available example of a UDG with kinematic data: AGC\,242019 \citep{Shi2021}. This galaxy is ideally inclined for kinematic work ($i=72^{\circ}\pm1^{\circ}$), as shown by both optical and \hi\ morphologies \citep{Shi2021}. High-resolution \hi\ data from the Very Large Array provides the rotation curve and gas surface density distribution, while Spitzer images at 3.6 $\mu$m trace the stellar mass distribution. \citet{Shi2021} conclude that the galaxy is inconsistent with MOND and has a cuspy DM halo, which is rather unusual for LSB dwarfs \citep[e.g.,][]{Li2020}. In the following, I present a new MOND analysis of the asymmetric-drift-corrected rotation curve from \citet{Shi2021}.

\citet{Shi2021} adopted a distance of $30.8\pm1.5$ Mpc from the Hubble flow, after correcting the \hi\ systemic velocity for Virgo, Great Attractor, and Shapley infall. An error of only 5$\%$ on an Hubble-flow distance is severely underestimated, considering that the accuracy of primary distance methods (Cepheids, supernovae, and the tip magnitude of the red giant branch, TRGB) is typically in the range $5-15\%$ \citep[e.g.,][]{Tully2013, Shaya2017, Schombert2020}. In addition, AGC\,242019 lies in the sky area of the Virgo galaxy cluster, where cosmic flow solutions can be triple-valued leading to significant ambiguities on flow distances \citep{Shaya2017}. Here, I adopt the same error scheme on flow distances used by the SPARC team: 25$\%$ for galaxies in the range $20 < D < 40$ Mpc \citep{Lelli2016a}. In addition to being more realistic, this distance uncertainty ensures that AGC\,242019 is analyzed using the same assumptions as SPARC galaxies \citep{Li2018}.

Figure\,\ref{fig:inclined} shows a MOND fit to the \hi\ rotation curve of AGC\,242019, using the same Bayesian method that \citet{Li2018} applied to the entire SPARC sample. Specifically, I ran a Markov chain Monte Carlo simulation in which three nuisance parameters are varied within their uncertainties: the inclination $i$, the distance $D$, and the stellar mass-to-light ratio $\Upsilon_\star$ \citep[see][for technical details]{Li2018}. I assumed Gaussian priors centered at $i=72^{\circ}\pm1^{\circ}$, $D=30.8\pm7.7$ Mpc, and $\Upsilon_\star=0.5\pm0.1$ M$_\odot$/L$_\odot$. In the case of AGC 242019, both $i$ and $\Upsilon_\star$ play virtually no role: the inclination is well constrained by the \hi\ data \citep{Shi2021}, while the stellar contribution is subdominant with respect to the gas contribution ($M_{\rm gas}/M_\star\simeq 8$, which is common in dwarf galaxies, see \citealt{Lelli2022}). The main role is played by galaxy distance, which is varied by about 2.5$\sigma$ from the prior center, giving $D=12.5\pm0.6$ Mpc. This is a significant variation but it is consistent with the flow model of \citet{Shaya2017}. At the sky position of AGC\,242019,  there are multiple solutions in Virgocentric flow models; for the systemic velocity of AGC\,242019 \citep[$\sim$1840 \kms,][]{Shi2021}, the low-distance branch gives a value close to 12.5 Mpc
\citep[see Fig.\,4 of][]{Shaya2017}. This distance agrees with the catalog of \citet{Castignani2022}, which classifies AGC\,242019 as a member of the Virgo-III filament that covers distances from $\sim$12 to $\sim$17 Mpc.

For consistency with the SPARC analysis, I did not consider uncertainties due to the disk thickness. A different disk thickness changes the amplitude of the baryonic contribution and the best-fit distance by $\pm$1.2 Mpc, which is consistent with the random errors within 2$\sigma$. In addition, for consistency with \citet{Li2018}, I used the MOND equation for isolated galaxies, neglecting the external field effect (EFE). The EFE in rotation curves is expected to be small and decreases the MOND boost only in the outermost points of the rotation curve \citep{Chae2020, Chae2021}. The EFE would improve the MOND fit in Fig.\,\ref{fig:inclined}, but at the cost of adding an extra free parameter (the external field strength).

If AGC\,242019 is at 12.5 Mpc rather than 30.8 Mpc, its half-light radius decreases from 3.7 kpc \citep{Shi2021} to 1.5 kpc, so the galaxy just fits into the UDG class. However, as discussed in Sect.\,\ref{sec:intro}, the UDG category is rather arbitrary and may have no reason to exist. More importantly, the predicted MOND distance can be tested using Hubble Space Telescope observations with the aim of detecting the TRGB; a significantly different distance would rule out MOND outright.

\begin{figure*}
\includegraphics[width=0.95\textwidth]{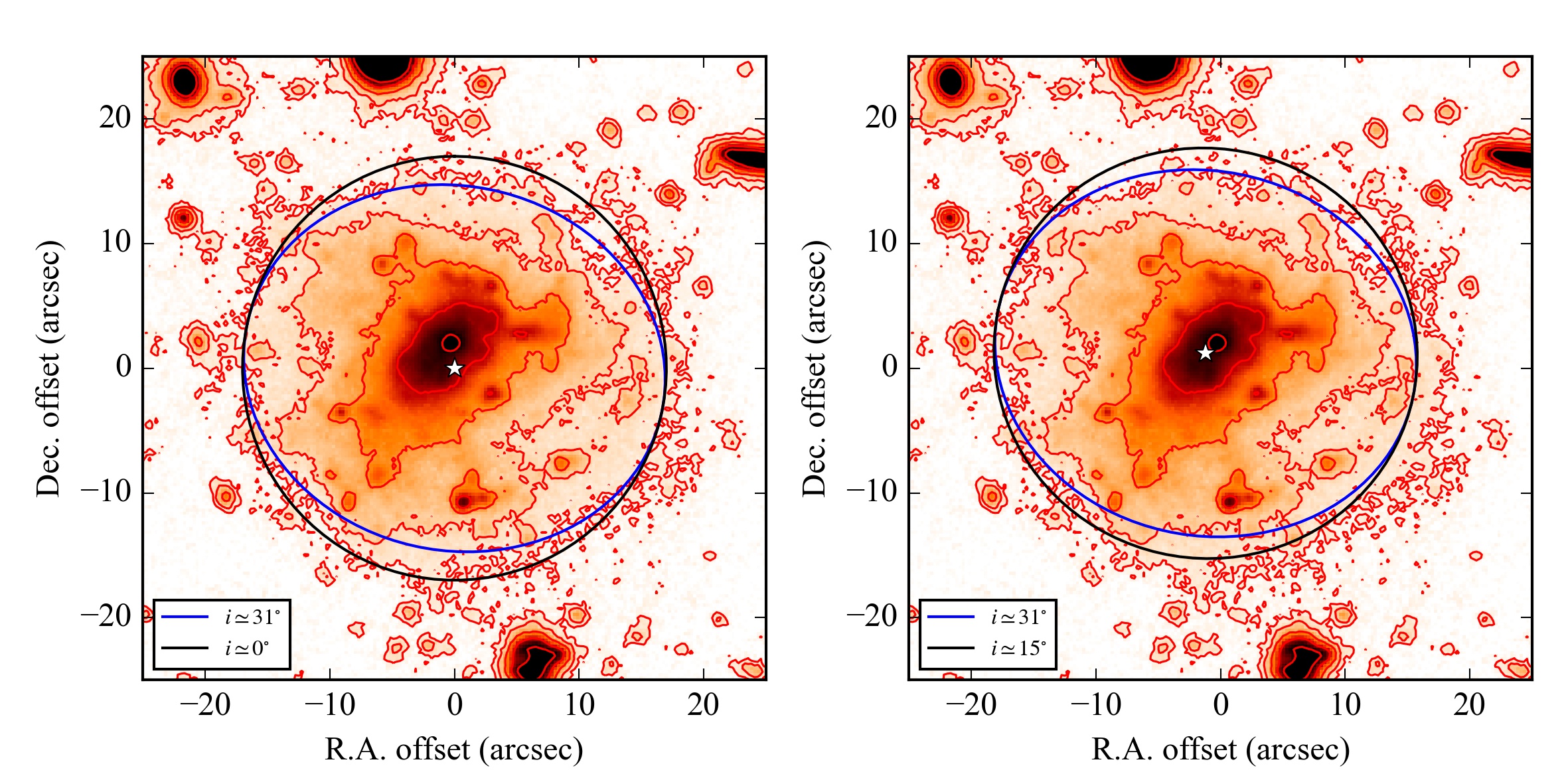}
\caption{$r$-band image from \citet{Mancera2024}. The red contours correspond to isophotes at 30, 29, 28, 27, and 26.3 mag arcsec$^{-2}$. In the left panel, the star shows the center from \citet{Mancera2024}; the blue ellipse has a position angle of $78^{\circ}$ and $b/a=0.86$ \citep[as in][]{Mancera2024}, while the black ellipse has $b/a=1$ (a perfectly face-on disk). In the right panel, the star is approximately at the center of the inner bar-like component; the blue ellipse has the same position angle and $b/a$ as in the left panel, while the black ellipse has $b/a\simeq0.97$ predicted by MOND (Fig.\,\ref{fig:faceon}). Given the irregular and asymmetric galaxy morphology, this axial ratio cannot be ruled out.}
\label{fig:isophotes}
\end{figure*}

\begin{figure*}
\includegraphics[width=0.45\textwidth]{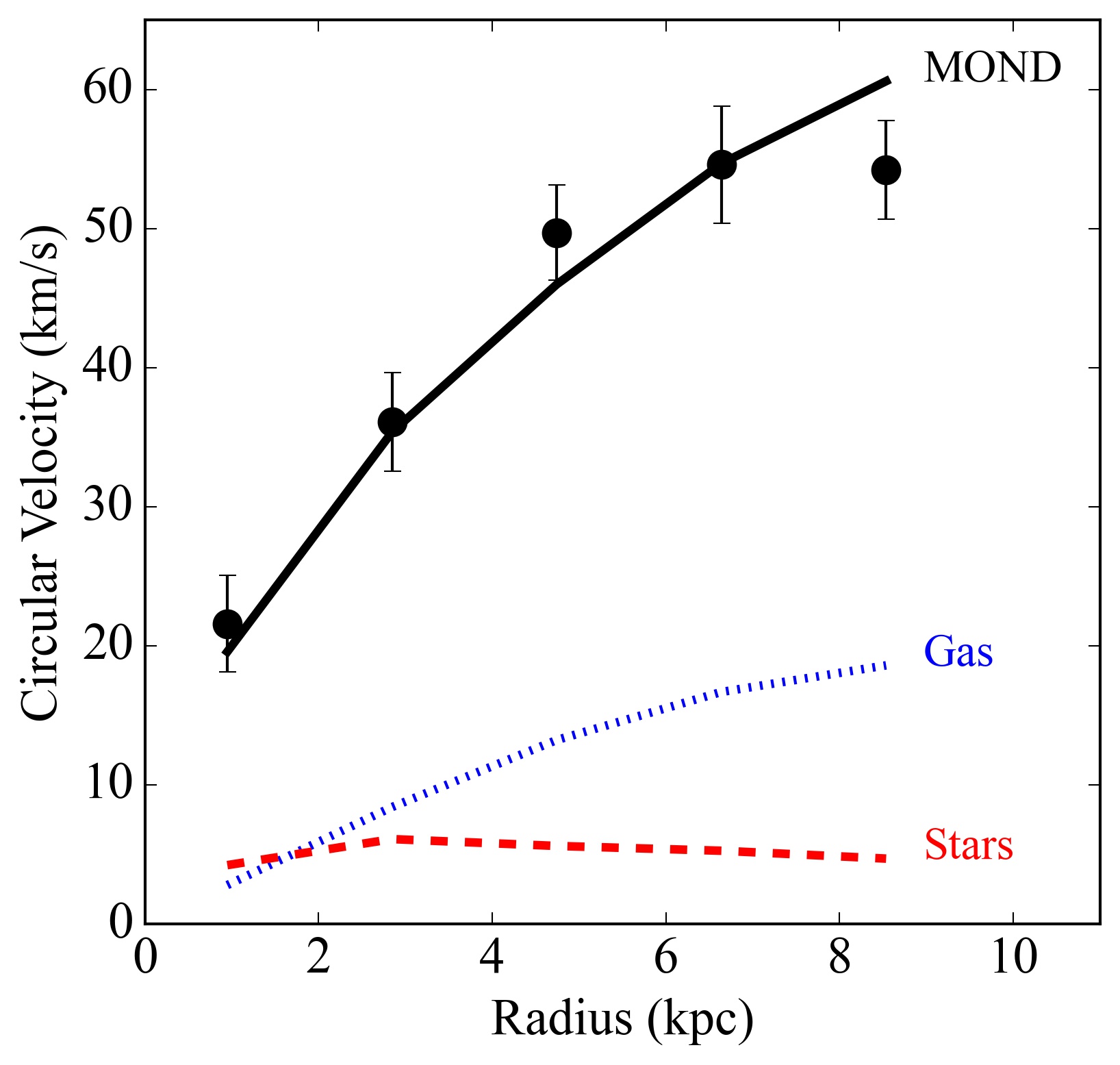}
\includegraphics[width=0.45\textwidth]{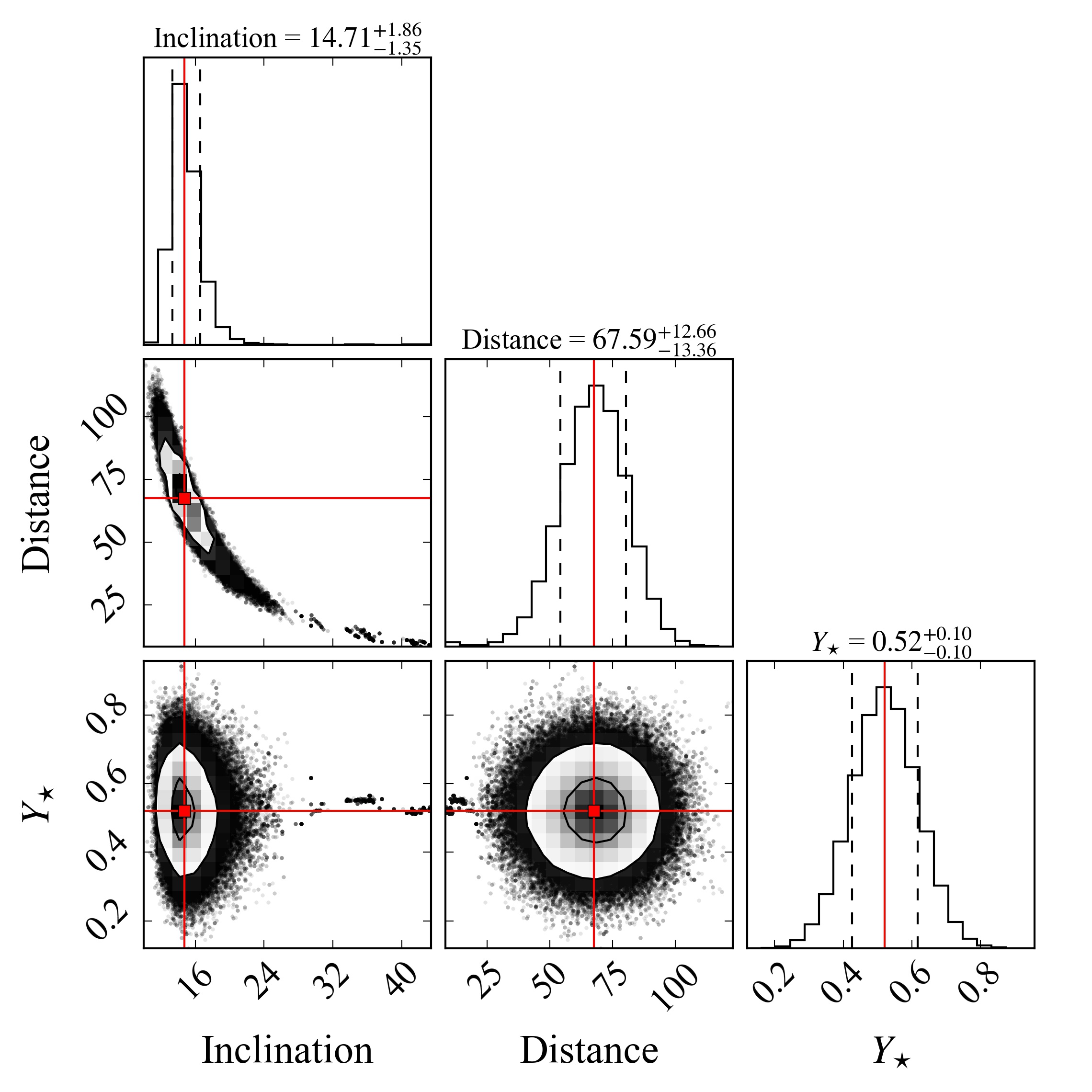}
\caption{Face-on ultra-diffuse galaxy AGC114905. The symbols and lines are the same as in Fig.\,\ref{fig:inclined}. MOND requires an inclination of about 15$^{\circ}$, which  is consistent with the optical image in Fig.\,\ref{fig:isophotes}.}
\label{fig:faceon}
\end{figure*}

\section{The face-on galaxy AGC 114905}\label{sec:pavel}

The face-on galaxy AGC\,114905 is studied in \citet{Mancera2022, Mancera2024}, who conclude that the galaxy rotation curve defies the predictions of both $\Lambda$CDM and MOND. Here I analyze the kinematic data from the most recent paper, which presents several improvements over the previous data. In particular, \citet{Mancera2024} obtained very deep optical images and measured $i=31^{\circ}\pm2^{\circ}$ by fitting the optical isophotes. In my experience, it is very hard to determine the inclination of face-on galaxies at this level of accuracy (two degrees), even for bright spiral galaxies, which  in principle should be much easier to study than dim LSB dwarfs. For example, the DiskMass survey \citep{Bershady2010} targeted bright face-on spirals (to measure their vertical velocity dispersion) and obtained kinematic inclinations from high-resolution H$\alpha$ velocity fields with a typical error of 5$^{\circ}$ \citep[see Sect. 5.1 in][]{Andersen2013}. The DiskMass survey  had to rely on the Tully-Fisher relation to measure robust inclinations with errors of 2$^{\circ}$ or less \citep{Andersen2013}. We are therefore in the puzzling situation in which face-on LSB dwarfs with lower quality data than bright spirals are used to argue for deviations from the BTFR (Sect.\,\ref{sec:inc}).

The $r$-band image from \citet{Mancera2024} is shown in Fig.\,\ref{fig:isophotes}. In the left panel, I consider the galaxy center from \citet{Mancera2024}, which does not correspond to the inner luminosity peak, but approximately corresponds to the center of the outermost isophotes. It is unclear whether the best-fit $b/a=0.86$ from \citet{Mancera2024} provides a better fit than $b/a=1$ (corresponding to a symmetric disk with $i=0^{\circ}$) because the light distribution is \emph{not} symmetric: $b/a=0.86$ fits well in the northern half, while $b/a=1$ fits well in the southern one. In the right panel, I consider the photometric center of the inner bar-like structure, shifting the previous center by eye by ($-1.2'',+1.2''$). Again, it seems impossible to accurately constrain the inclination because the outer asymmetry is enhanced. Another choice of the center may be the peak luminosity ($\sim$26.3 $r$ mag\,arc$^{-1}$) at approximately ($-0.3'', 2.0''$), but it is unclear whether this is the galaxy nucleus, a star-forming region, or some other object projected along the line of sight. In any case, either the light distribution is strongly asymmetric in the inner parts (similarly to the offset bar in the LMC) or strongly lopsided in the outer parts. An outer lopsided distribution has actually been predicted by MOND, due to the EFE \citep{Banik2022b}. Considering all of the above, I   adopt $i=31^{\circ}$, as in \citet{Mancera2024}, but with a more realistic uncertainty of $\pm5^\circ$. This uncertainty remains conservatively low because $i=0^{\circ}$ would represent a more than 6$\sigma$ deviation, which is arguably not the case (Fig.\,\ref{fig:isophotes}). The true errors on $i$ cannot be symmetric: we can rule out at high confidence ($>$3$\sigma$) a disk with $i\gtrsim45^{\circ}$, but the same is not true for a face-one disk with $i\lesssim15^{\circ}$.

Regarding the distance, \citet{Mancera2024} assumed $D=78.7\pm1.5$ Mpc from the Hubble flow. An error of just 2$\%$ on a Hubble-flow distance is clearly underestimated because it is smaller than the current uncertainties on the Hubble constant \citep[e.g.,][]{Tully2023}. At the redshift of AGC\,114905, however, peculiar velocities start to become subdominant with respect to the recession velocities, so the relative distance error is smaller than for galaxies at lower redshifts. Following again the SPARC error scheme \citep{Lelli2016a}, I  adopt $D=78.7\pm11.8$ Mpc.

Figure\,\ref{fig:faceon} shows the  MOND fits to the \hi\ rotation curve, using the same method as in Sect.\,\ref{sec:shi}. The Gaussian priors are centered at $i=31^{\circ}\pm5^{\circ}$, $D=78.7\pm11.8$ Mpc, and $\Upsilon_\star=1\pm0.35$. The value of $\Upsilon_\star$ does not correspond to a specific photometric band,  but is just meant to allow for uncertainties of about 0.15 dex around the stellar contribution estimated by \citet{Mancera2024}, who used color profiles to determine a radially varying $\Upsilon_\star$. In any case, the value of $\Upsilon_\star$ plays nearly no role because the galaxy is heavily gas dominated ($M_{\rm gas}/M_{\star}\simeq10$). MOND can fit the observed rotation curve with $i=15^{\circ}$ (about 3$\sigma$ from the prior center) and $D=67.6$ Mpc (about 1$\sigma$ from the prior center). Our MOND-inferred inclination is slightly higher than the similar measurement of 11$^{\circ}$ from \citet{Mancera2024} because $i$ and $D$ are degenerate: one can decrease $i$ while increasing $D$, or vice versa. In any case, both inclination values are clearly consistent with the optical image (Fig.\,\ref{fig:isophotes}), considering the irregular, asymmetric isophotes and the ambiguities in the galaxy center.

\section{Conclusions}

In this letter I showed that gas-rich UDGs and gas-rich LSB galaxies lie on the same BTFR when one considers data with similarly low quality (face-on and/or asymmetric disks). Observational uncertainties  systematically scatter galaxies toward  low rotation velocities. I also showed that the two gas-rich UDGs with the best available rotation-curve data (AGC\,242019 and AGC\,114905) are consistent with MOND when one considers realistic uncertainties on inclination and distance.

Underestimated uncertainties on distances and inclinations affect previous conclusions on the DM halos of these galaxies, such as the presence of a cuspy DM halo in AGC\,242019 \citep{Shi2021} and a severe DM deficiency in AGC\,114905 \citep{Mancera2022, Mancera2024}. Addressing these issues is beyond the scope of this letter. In general, however, UDGs do not seem to have a peculiar dynamical behavior compared to gas-rich LSB dwarfs from SPARC \citep{Lelli2016a}, so it seems natural to expect that their DM halos are similar to those of other galaxies in the $\Lambda$CDM context \citep{Li2020}.

\begin{acknowledgements}
I thank Pavel Mancera-Pi$\tilde{\rm{n}}$a and Yong Shi for providing the data of AGC\,114905 and AGC\,242019, respectively. I also thank Brent Tully for clarifications on triple-valued cosmic-flow distances. Finally, I thank the organizers of the workshop ``A new dawn of dwarf galaxies'' (Oliver M\"uller, Katja Fahrion, Teymoor Saifollahi, Maria Angela Raj, and Noam Libeskind) at the Lorentz Center (Leiden, The Netherlands) where some of these issues were discussed, pushing me to have an in-depth look into UDGs.
\end{acknowledgements}

%
%

\bibliographystyle{aa} 
\bibliography{GasDynamics} 
\end{document}